\documentstyle[12pt]{article}
\begin{document}

\title{LEVEL CROSSING ALONG SPHALERON BARRIERS}
\vspace{1.5truecm}
\author{
{\bf Jutta Kunz}\\
Fachbereich Physik, Universit\"at Oldenburg, Postfach 2503\\
D-26111 Oldenburg, Germany\\
and\\
Instituut voor Theoretische Fysica, Rijksuniversiteit te Utrecht\\
NL-3508 TA Utrecht, The Netherlands
\and
{\bf Yves Brihaye}\\
Facult\'e des Sciences, Universit\'e de Mons-Hainaut\\
B-7000 Mons, Belgium}

\vspace{1.5truecm}

\date{March 1, 1994}

\maketitle
\vspace{1.0truecm}

\begin{abstract}

In the electroweak sector of the standard model
topologically inequivalent vacua are separated by finite
energy barriers, whose height is given by the sphale\-ron.
For large values of the Higgs mass
there exist several sphaleron solutions
and the barriers are no longer symmetric.
We construct paths of classical configurations
from one vacuum to a neighbouring one
and solve the fermion equations
in the background field configurations along such paths,
choosing the fermions of a doublet degenerate in mass.
As in the case of light Higgs masses we observe
the level crossing phenomenon also for large Higgs masses.

\end{abstract}
\vfill\eject

\section{Introduction}

In 1976 't Hooft [1] observed that
the standard model does not absolutely conserve
baryon and lepton number
due to the Adler-Bell-Jackiw anomaly.
In particular 't Hooft considered
spontaneous fermion number violation due to instanton
induced transitions, finding a negligible amplitude for
such processes.
At high energies, however, fermion number violating
tunnelling transitions between
topologically distinct vacua might
become observable at future accelerators [2,3].

Manton [4] considered
the possibility of fermion number
violation in the standard model
from another point of view.
Investigating the topological structure
of the configuration space of the Weinberg-Salam theory,
Manton showed that there are non-contractible
loops in configuration space, and
predicted the existence of a static, unstable solution
of the field equations,
a sphaleron with Chern-Simons charge $N_{\rm CS} = 1/2$ [5],
representing the top of the energy barrier between
topologically distinct vacua.

At finite temperature this energy barrier between
topologically distinct vacua can be overcome
due to thermal fluctuations of the fields,
and fermion number violating
vacuum to vacuum transitions
involving changes of baryon and lepton number
can occur.
Baryon number violation in the standard model due to
thermal transitions over the barrier may be relevant
for the generation of the baryon asymmetry of the universe
[6-10].

The change of fermion number during a vacuum to vacuum
transition over the barrier is associated with
the level crossing of one fermionic mode.
A fermionic eigenstate
emerges from the continuum of positive energy states, crosses
zero energy and dives into the Dirac sea of negative
energy states.

While it has been known since some time, that the fermions possess
a zero mode in the background field of the sphaleron [11-13],
the level crossing phenomenon in the
background field of the sphaleron barrier
was only demonstrated recently,
for fermion doublets degenerate in mass,
moderate values of the Higgs mass
and vanishing mixing angle [14-16].

For large values of the Higgs mass
the barrier separating topologically distinct vacua
becomes asymmetric, the top of the barrier being now
represented by a bisphaleron with Chern-Simons charge
$N_{\rm CS} \ne 1/2$ [17,18].
Here we demonstrate the level crossing phenomenon
for fermions in the background field of
such asymmetric barriers for large values of the Higgs mass.
We numerically determine the fermion eigenvalues along
various paths from one vacuum to another, passing a bisphaleron.
In particular we consider the extremal energy path [19],
which however becomes unsatisfactory for large values of the
Higgs mass [20], and construct improved paths.
We again assume that the fermions of a doublet are degenerate in mass.
This assumption, violated in the standard model,
allows for spherically symmetric ans\"atze
for all of the fields, when the mixing angle dependence
is neglected (which is an excellent approximation [21,22]).

We briefly review in section 2
the Weinberg-Salam Lagrangian with the approximations employed.
In section 3 we discuss paths over the barrier for large values of the
Higgs mass,
providing the background fields for the fermions.
In section 4 we derive the radial equations for
the fermions and present our results for the fermion eigenvalues
along the paths.
We give our conclusions in section 5.

\section{\bf Weinberg-Salam Lagrangian}

Let us consider the bosonic sector of the Weinberg-Salam theory
in the limit of vanishing mixing angle.
In this limit the U(1) field
decouples and can consistently be set to zero.
\begin{equation}
{\cal L}_{\rm b} = -\frac{1}{4} F_{\mu\nu}^a F^{\mu\nu,a}
+ (D_\mu \Phi)^{\dagger} (D^\mu \Phi)
- \lambda (\Phi^{\dagger} \Phi - \frac{v^2}{2} )^2
\   \end{equation}
with the SU(2)$_{\rm L}$ field strength tensor
\begin{equation}
F_{\mu\nu}^a=\partial_\mu V_\nu^a-\partial_\nu V_\mu^a
            + g \epsilon^{abc} V_\mu^b V_\nu^c
\ , \end{equation}
and the covariant derivative for the Higgs field
\begin{equation}
D_{\mu} \Phi = \Bigl(\partial_{\mu}
             -\frac{i}{2}g \tau^a V_{\mu}^a \Bigr)\Phi
\ . \end{equation}
The ${\rm SU(2)_L}$
gauge symmetry is spontaneously broken
due to the non-vanishing vacuum expectation
value $v$ of the Higgs field
\begin{equation}
    \langle \Phi \rangle = \frac{v}{\sqrt2}
    \left( \begin{array}{c} 0\\1  \end{array} \right)
\ , \end{equation}
leading to the boson masses
\begin{equation}
    M_W = M_Z =\frac{1}{2} g v \ , \ \ \ \ \ \
    M_H = v \sqrt{2 \lambda}
\ . \end{equation}
We employ the values $M_W=80 {\rm GeV}$, $g=0.65$.

For vanishing mixing angle,
considering only fermion doublets degenerate in mass,
the fermion Lagrangian reads
\begin{eqnarray}
{\cal L}_{\rm f} & = &
   \bar q_{\rm L} i \gamma^\mu D_\mu q_{\rm L}
 + \bar q_{\rm R} i \gamma^\mu \partial_\mu q_{\rm R}
   \nonumber \\
           & - & f^{(q)} \bar q_{\rm L}
           (\tilde \Phi u_{\rm R} + \Phi d_{\rm R})
               - f^{(q)} (\bar d_{\rm R} \Phi^\dagger
                     +\bar u_{\rm R} \tilde \Phi^\dagger)
           q_{\rm L}
\ , \end{eqnarray}
where $q_{\rm L}$ denotes the lefthanded doublet
$(u_{\rm L},d_{\rm L})$,
while $q_{\rm R}$ abbreviates the righthanded singlets
$(u_{\rm R},d_{\rm R})$,
with covariant derivative
\begin{equation}
D_\mu q_{\rm L} = \Bigl(\partial_{\mu}
             -\frac{i}{2}g \tau^a V_{\mu}^a \Bigr) q_{\rm L}
\ , \end{equation}
and with $\tilde \Phi = i \tau_2 \Phi^*$.
The fermion mass is given by
\begin{equation}
M_F=\frac{1}{\sqrt{2}}f^{(q)} v
\ . \end{equation}

All gauge field configurations can be classified by a charge,
the Chern-Simons charge.
The Chern-Simons current
\begin{equation}
 K_\mu=\frac{g^2}{16\pi^2}\varepsilon_{\mu\nu\rho\sigma} {\rm Tr}(
 {\cal F}^{\nu\rho}
 {\cal V}^\sigma
 + \frac{2}{3} i g {\cal V}^\nu {\cal V}^\rho {\cal V}^\sigma )
\   \end{equation}
(${\cal F}_{\nu\rho} = 1/2 \tau^i F^i_{\nu\rho}$,
${\cal V}_\sigma = 1/2 \tau^i V^i_\sigma$)
is not conserved,
its divergence $\partial^\mu K_\mu$
represents the U(1) anomaly.
The Chern-Simons charge
of a configuration is given by
\begin{equation}
N_{\rm CS} = \int d^3r K^0
\ . \end{equation}
For the vacua the Chern-Simons charge is identical to the
integer winding number.

\section{\bf Classical bosonic configurations}

Topologically inequivalent vacua are separated by finite
energy barriers, whose height is determined by
an unstable classical solution.
For small values of the Higgs mass this is the sphaleron [5],
for large values of the Higgs mass the unstable solution
with only one negative mode is the first bisphaleron [17,18].

\subsection{\bf Spherically symmetric ansatz}

In the limit of vanishing mixing angle the sphaleron and bisphalerons
are spherically symmetric.
The general static, spherically symmetric ansatz for the
gauge and Higgs fields is given by [23]
\begin{eqnarray}
    \Phi & = & \frac{v}{\sqrt {2}}
  \Bigl(H(r) + i \vec \tau \cdot \hat r K(r)\Bigr)
    \left( \begin{array}{c} 0\\1  \end{array} \right)
  \ , \\
  V_i^a & = & \frac{1-f_A(r)}{gr} \epsilon_{aij}\hat r_j
  + \frac{f_B(r)}{gr} (\delta_{ia}-\hat r_i \hat r_a)
  + \frac{f_C(r)}{gr} \hat r_i \hat r_a  \ , \\
  V_0^a & = & 0
\ , \end{eqnarray}
and involves the five radial functions $H(r)$, $K(r)$,
$f_A(r)$, $f_B(r)$ and $f_C(r)$.

This ansatz leads to the energy functional
\begin{eqnarray}
 E & = & \frac{4\pi M_W}{g^2} \int^{\infty}_0 dx
         \Bigl[  \frac{1}{2x^2} (f^2_A + f^2_B  -1)^2
         + (f'_A + \frac{f_Bf_C}{x})^2
         + (f'_B - \frac{f_Af_C}{x})^2
\nonumber \\
   & + & (K^2+H^2) (1+f_A^2+f^2_B +
         \frac{f_C^2}{2})+2f_A (K^2-H^2) - 4f_B H K
\nonumber \\
   & + & 2x^2(H'^2+K'^2) - 2xf_C (K'H - KH') +
         \epsilon x^2 (H^2+K^2 -1)^2
    \Bigr]
\ , \end{eqnarray}
where $x=M_Wr$, the prime means derivative with respect to $x$,
and
\begin{equation}
\epsilon = {{4\lambda} \over g^2} = {1\over 2} ({M_H \over M_W})^2 \ .
\end{equation}

\par The spherically symmetric ansatz
 does not completely break the gauge symmetry,
there is a residual invariance of the form
\begin{equation}
(f_A+if_B)\longrightarrow e^{i\theta} (f_A+if_B) \ ,
\end{equation}
\begin{equation}
(H+iK)\longrightarrow e^{i{\theta\over 2}} (H+iK) \ ,
\end{equation}
\begin{equation}
f_C \longrightarrow f_C+x\theta' \ .
\end{equation}
In order to construct classical solutions,
the residual gauge freedom has to be fixed.
Here we choose the radial gauge  $(x_iV_i = 0)$, imposing
\begin{equation}
f_C(x) = 0 \ .
\end{equation}
This still leaves a global ${\rm U}(1)$ symmetry,
which is fixed by imposing the conditions
\begin{eqnarray}
&f_A(0) = 1\ ,\ f_B(0) =0 \ , \\
&f_A(\infty) = \cos 2\pi q\ ,\ f_B(\infty) = \sin 2\pi q\ , \\
&H(\infty) = \cos \pi q\ ,\ K(\infty) = \sin \pi q \ ,
\end{eqnarray}
and $q\in[0,1]$.

\par The spherically symmetric ansatz yields
the Chern-Simons charge
\begin{equation}
N_{\rm CS} = q+{1\over {2\pi}} \int^{\infty}_0 dx(f'_Af_B-f_Af'_B) \ .
\end{equation}

\subsection{\bf Classical solutions}

In the following
we denote a generic configuration of the bosonic fields by
\begin{equation}
C = (f_A, f_B, H, K) \ .
\end{equation}
We characterize a finite energy configuration further by a quadruple
of numbers
\begin{equation}
X(C) = \lbrace H(0), q, E, N_{\rm CS} \rbrace \ .
\end{equation}
Defining $P$ as the parity operator supplemented by a
gauge transformation Eqs.~(16)-(18) with $\theta=2\pi$,
\begin{equation}
P(f_A,f_B,,H,K) = (f_A,-f_B,-H,K) \ \ \ , \ \
X(P(C)) = \lbrace -H(0),1-q,E,1-N_{\rm CS} \rbrace \ .
\end{equation}
This choice of $P$
preserves Eqs.~(20)-(22).

We are left with two (artificially)
distinct vacuum configurations
\begin{equation}
C_{\rm vac} \equiv  V_{\pm} = (1,0,\pm 1,0) \quad , \quad
X(V_+) = \lbrace 1,0,0,0 \rbrace \quad , \quad
X(V_-) = \lbrace -1,1,0,1 \rbrace \ .
\end{equation}

For $\epsilon < 72$ (i.~e.~$M_H<12 M_W$) only the sphaleron solution [5]
is known.
The sphaleron solution is invariant
under the operator $P$
\begin{equation}
C_{\rm sp} = (f_A, 0 , 0 , K) \quad ,
\quad X(C_{\rm sp}) =
\lbrace 0, {1\over 2}, E(\epsilon), {1\over 2} \rbrace
\ . \end{equation}
The non-trivial functions $f_A$ and $K$
are shown in Fig.~1 for several values of the Higgs mass.
The classical energy increases monotonically as a function of the Higgs mass,
e.~g.~$E(\epsilon = 0)\simeq 7.23\ {\rm TeV}$,
$E(\epsilon = 100) = 11.76 \ {\rm TeV}$,
$E(\epsilon = 200) = 11.99 \ {\rm TeV}$,
$E(\epsilon = \infty) = 12.87\ {\rm TeV}$.

For $\epsilon > 72$ further classical solutions exist [17,18],
the bisphalerons, for which all four functions $f_A,f_B,H$ and $K$
are non-trivial. They are shown in Fig.~2.
Their energy and Chern-Simons charge both depend on $M_H$, e.~g.
\begin{eqnarray}
X(C_{\rm bi})(\epsilon = 100)
  &=& \lbrace 0.53, 0.53, 11.69, 0.563 \rbrace \ ,\\
X(C_{\rm bi})(\epsilon = 200)
  &=& \lbrace 0.80, 0.55, 11.88, 0.597 \rbrace \ ,\\
X(C_{\rm bi})(\epsilon = \infty)
  &=& \lbrace 1.00, 0.56, 12.06, 0.625 \rbrace \ ,
\end{eqnarray}
where the energy values $E(\epsilon)$ are given in TeV.
In contrast to the sphaleron Eq.~(28), the bisphalerons are not invariant
under $P$ but occur as parity doublets.
All bisphalerons are lower in energy than the sphaleron.
Therefore the first bisphalerons represent the top of the energy
barrier between inequivalent vacua for $M_H>12 M_W$.

\subsection{\bf Unstable modes}

To obtain the unstable modes about the sphaleron and the
bisphalerons,
one considers a general spherically symmetric fluctuation
about the classical solutions.
Insertion of the configuration
\begin{equation}
(f_A,f_B,H,K,f_C) = (C_{\rm cl},0)
+ (\eta_A, \eta_B, \eta_H, \eta_K, \eta_C)
\end{equation}
into the energy functional $E$
then leads to an eigenvalue equation for the discrete modes.
For $\epsilon < 72$ the sphaleron has exactly one unstable mode,
while for $\epsilon > 72$
the sphaleron has at least two unstable modes [18,20].
The first bisphalerons possess only one unstable mode [18,24].

The negative eigenmode gives the direction of instability.
A simple calculation involving the unstable mode
of the first bisphalerons
therefore gives a rough impression
of the energy barrier in the vicinity of the bisphalerons.
To this end we consider the configurations
\begin{equation}
(f_A,f_B,H,K,f_C) = (C_{\rm bi},0)
+ \lambda(\eta_A, \eta_B, \eta_H, \eta_K, \eta_C) \ ,
\end{equation}
where $\lambda$ is a multiplicative constant.
The fluctuation eigenvector $\eta$ for the single unstable mode
of the first bisphaleron is
presented in Fig.~3 for $\epsilon =100$
(with the normalisation $\eta_H(0)=1$).
Gauging rotating the configurations
Eq.~(33) into the radial gauge Eqs.~(19)-(22) we find
\begin{equation}
H(0) = H_{\rm bi}(0) + \lambda \ ,
 \quad q(\lambda) = q_{\rm bi} - c \lambda \ ,
 \quad c = {1\over 2\pi} \int_0 ^{\infty} \frac{\eta_C}{x} dx \ .
\end{equation}

The configurations Eq.~(33) are regular and have finite energy.
The energy of these configurations decreases
as a function of $\lambda$ to both sides of
the bisphaleron (where $\lambda = 0$)
up to finite values $\lambda_{\pm}$.
For example for $\epsilon = 100$, $c = 0.085$ and
\begin{eqnarray}
  \lambda_- = \ \ 0.42\ , \quad
X &=& \lbrace 0.95 , 0.492 , 11.56 , 0.518 \rbrace \ ,\\
  \lambda_+ = -1.00\ , \quad
X &=& \lbrace -0.47 , 0.615 , 10.87 , 0.666 \rbrace \ ,
\end{eqnarray}
to be compared with the bisphaleron values ($\lambda=0$), Eq.~(29).

\subsection{\bf Vacuum to vacuum paths}

Let us now consider vacuum to vacuum paths across the finite
energy barriers. Such paths are also called
non-contractible loops (NCLs) [4].
In the gauge used, a NCL represents
a path in the space of finite energy configurations
between the vacua $V_+$ and $V_-$.
Along such a path the asymptotic angle $2\pi q$
of the functions $f_A$ and $f_B$ covers the trigonometric circle,
while the boundary conditions at the origin
of the functions $f_A$ and $f_B$, Eq.~(20), are maintained.

The non-contractible loop constructed by Manton [4]
(see also [25]) is symmetric under parity and its
energy culminates at the sphaleron.
It is therefore not appropriate for large Higgs masses,
where the first bisphaleron represents the top of the energy barrier.
A loop reaching its maximal energy at the bisphaleron
was constructed by Klinkhamer [25].

A systematic technique to numerically construct
an extremal path over the barrier
is provided by the functional
\begin{equation}
 W = E+\xi \ {8\pi M_W\over {g^2}} \int dx (f'_Af_B-f'_Bf_A)
\ , \end{equation}
where the Chern-Simons charge is added to the energy functional
by means of the Lagrange multiplier $\xi$ [19].
For $\xi=0$ this functional yields the sphaleron and the bisphalerons,
while for $\xi \rightarrow \pm 1$ the extrema of $W$ approach
the vacua $V_{\pm}$, independently of $M_H$ [20].
The extremal energy path obtained from $W$ is invariant
under parity $P$.

For small values of the Higgs mass (up to $M_H<12M_W$)
the functional $W$ leads
to a symmetric smooth minimal energy path over the barrier
with the sphaleron at the top at $N_{\rm CS}=1/2$.
However, for large values of the Higgs mass ($M_H>12M_W$)
the extremal energy path obtained from the functional $W$
fails to be satisfactory
in the neighborhood of the bisphaleron,
due to the occurrence of a catastrophe [20].
As demonstrated in Figs.~4a and 4b,
where we show the energy as a function of the Chern-Simons
number along the extremal path,
a spike occurs in the vicinity of the bisphaleron.
The path coming from the vacuum $V_+$
does not culminate at the bisphaleron,
but increases in energy up to the sphaleron,
where it continues symmetrically
to the parity conjugate bisphaleron
to finally reach the vacuum $V_-$.
In Fig.~4a we show
the complete extremal energy path for $\epsilon =200$
and compare with the minimal energy path for $\epsilon =0.5$.
In Fig.~4b we show the energy as a function of the Chern-Simons
number only in the vicinity of the bisphaleron for $\epsilon =100$.
Here we additionally present the energy
for the configurations Eq.~(33) involving the negative mode.
The plot suggests that these configurations can be used
to improve the extremal path in the region of the bisphaleron.

We therefore now consider another construction of a NCL,
reaching its maximal energy value at the bisphaleron.
This NCL makes use of the invariance of the
energy density under $P$ as well as
of the unique negative eigenmode of the bisphaleron.
Far enough away from the bisphaleron,
the path coincides with the extremal energy path,
approaching the vacua on both sides.
While in the vicinity of the bisphaleron the path
is altered to reach its maximal energy at the bisphaleron
and to prevent the occurrence of the spike.

Let us call this path $L(N_{\rm CS})$
and consider it in detail for $\epsilon = 100$.
We denote the two sets of extrema
reaching the vacua $V_+$ and $V_-$
as $C_{\rm e}^+(N_{\rm CS})$ and $C_{\rm e}^-(N_{\rm CS})$,
respectively.
In the vicinity of the vacua $V_{\pm}$ we then choose
\begin{eqnarray}
L(N_{\rm CS}) = C_{\rm e}^+(N_{\rm CS}) \
   {\rm from} \   N_{\rm CS} &= 0    \
  &{\rm i.e.} \  X=\lbrace 1,0,0,0 \rbrace \\
   {\rm to}   \   N_{\rm CS} &= 0.54 \
  &{\rm i.e.} \  X=\lbrace 0.89, 0.50, 11.52, 0.54 \rbrace \\
L(N_{\rm CS}) = C_{\rm e}^-(N_{\rm CS}) \
   {\rm from} \   N_{\rm CS} &= 0.62 \
  &{\rm i.e.} \  X=\lbrace -0.98, 0.61, 9.35, 0.62 \rbrace \\
   {\rm to}   \   N_{\rm CS} &= 1 \
  &{\rm i.e.} \  X= \lbrace -1,1,0,1 \rbrace
\ . \end{eqnarray}
In the vicinity of the bisphaleron
($N_{\rm CS,bi} = 0.563$),
i.~e.~from $N_{\rm CS}= 0.54$ to $0.62$,
the loop is closed by a set of suitable linear combinations of
extrema of $W$ and of configurations Eq.~(33),
having the same asymptotic value of $q$
\begin{equation}
L(N_{\rm CS}) = \Bigl(1 - \beta (q)\Bigr)
                C^{\sigma}_{\rm e}(N_{\rm CS})
           +  \beta (q) \Bigl( C_{\rm bi} + {{q_{\rm bi} -q}\over c}
            (\eta_A, \eta_B, \eta_C, \eta_H, \eta_K)
             \Bigr)_{\theta(q)}
\ , \end{equation}
\begin{equation}
q=q(N_{\rm CS}) \ , \quad
 0 \leq \beta (q) \leq 1 \ , \quad
 \sigma = {\rm sign}(q_{\rm bi}-q)
\ , \end{equation}
where $q$ corresponds to the asymptotic angle
of the extremum $C_{\rm e}^{\sigma}(N_{\rm CS})$
(the coefficient in front of $\eta$ is chosen accordingly
(see Eq.~(33)),
and the subscript $\theta(q)$
denotes that the configurations are gauge rotated to the radial gauge.
Note, that it is essential to combine
configurations with identical asymptotic angle $q$
in order to keep the energy finite.
The configurations Eqs.~(42)-(43) have to coincide with
the extremal configurations Eqs.~(39) and (40),
when $N_{\rm CS}$ = 0.54 and 0.62, respectively,
and with the bisphaleron for
$N_{\rm CS}= 0.563$.
Therefore the function $\beta (q)$
has to obey the conditions
\begin{equation}
\beta (q_{\rm bi}) = 1 \ , \quad   q_{\rm bi} = 0.53
\ , \end{equation}
\begin{equation}
\beta (q = 0.50) = 0 \ , \quad \beta (q = 0.61) = 0
\ . \end{equation}
Apart from these constraints, the function $\beta$
should increase (resp.~decrease) monotonically
for $q \leq q_{\rm bi}$ (resp.~$q \geq q_{\rm bi}$).
The smoothness of the NCL depends on the smoothness of $\beta$.
There remains, however, a large arbitrariness in the choice of
the function $\beta(q)$ as well as in the points chosen
to depart from the extremal path.
For a particular choice we show in Fig.~4b
the energy along such a loop as a function of $N_{\rm CS}$.
When departing from the left branch of the extremal path,
$C_{\rm e}^+(N_{\rm CS})$,
the energy increases monotonically up to the bisphaleron.
Beyond the bisphaleron the energy decreases monotonically,
until it reaches the right branch of the extremal path,
$C_{\rm e}^-(N_{\rm CS})$.
Thus this full vacuum to vacuuum path
is smooth and asymmetric, culminating at the bisphaleron.
Of course,
the parity conjugate path passes through the parity
conjugate bisphaleron.

Somewhere along the NCL the Higgs field must vanish at the origin
[4,19].
Considering the complete path Eqs.~(38)-(43),
this happens at a point of the part Eqs.~(42)-(43) of the path.
There the function $H(x)$ changes sign at the origin.
(In contrast to the loops of Refs.~[4,19]
the function $H(x)$ remains non-trivial along the whole path.)

\section{\bf Fermions}

We now derive and solve the fermion equations
in the background field configurations of
the vacuum to vacuum paths considered above.
To retain spherical symmetry
we consider only fermion doublets degenerate in mass.

\subsection{\bf Fermion equations}

{}From the fermion Lagrangian (6)
we obtain the eigenvalue equations
for the lefthanded doublet
\begin{equation}
i D_0 q_{\rm L} + i \sigma^i D_i q_{\rm L}
-f^{(q)} (\tilde \Phi u_{\rm R} + \Phi d_{\rm R} )=0
\ , \end{equation}
and for the righthanded singlets
\begin{equation}
i \partial_0
\left( \begin{array}{c} u_{\rm R}\\d_{\rm R} \end{array} \right)
-i \sigma^i \partial_i
\left( \begin{array}{c} u_{\rm R}\\d_{\rm R} \end{array} \right)
-f^{(q)}
\left( \begin{array}{c} \tilde \Phi^\dagger q_{\rm L}\\
             \Phi^\dagger q_{\rm L} \end{array} \right)
             =0
\ . \end{equation}

We employ the spherically symmetric ansatz for
the fermion eigenstates, the hedgehog ansatz,
\begin{equation}
q_{\rm L}(\vec r\,,t) = e^{-i\omega t}
\bigl( G_{\rm L}(r)
+ i \vec \sigma \cdot \hat r F_{\rm L}(r) \bigr) \chi_{\rm h}
\ , \end{equation}
\begin{equation}
q_{\rm R}(\vec r\,,t) = e^{-i\omega t}
\bigl( G_{\rm R}(r)
+ i \vec \sigma \cdot \hat r F_{\rm R}(r) \bigr) \chi_{\rm h}
\ , \end{equation}
with the hedgehog spinor
satisfying the spin-isospin relation
$\vec \sigma \chi_{\rm h} + \vec \tau \chi_{\rm h} = 0 $.

The left radial functions $G_{\rm L}$ and $F_{\rm L}$
transform under the residual gauge transformation Eqs.(16)-(18) as
\begin{equation}
(G_{\rm L}+iF_{\rm L})\longrightarrow
  e^{-i\frac{\theta}{2}} (G_{\rm L}+iF_{\rm L})
\ , \end{equation}
while the right radial functions $G_{\rm R}$ and $F_{\rm R}$
are invariant.

In the radial gauge, we obtain the following set of four
coupled first order differential equations [14]
\begin{equation}
\tilde \omega G_{\rm L} - F'_{\rm L} - \frac{2}{x}F_{\rm L}
+\frac{1-f_A}{x} F_{\rm L}
-\frac{f_B}{x} G_{\rm L}
-\frac{f_C}{2x}G_{\rm L}
-\tilde M_F(H G_{\rm R} + K F_{\rm R}) = 0
\ , \end{equation}
\begin{equation}
\tilde \omega F_{\rm L} + G'_{\rm L}
+\frac{1-f_A}{x} G_{\rm L}
+\frac{f_B}{x} F_{\rm L}
-\frac{f_C}{2x}F_{\rm L}
-\tilde M_F(H F_{\rm R} - K G_{\rm R}) = 0
\ , \end{equation}
\begin{equation}
\tilde \omega G_{\rm R} + F'_{\rm R} + \frac{2}{x}F_{\rm R}
-\tilde M_F(H G_{\rm L} - K F_{\rm L}) = 0
\ , \end{equation}
\begin{equation}
\tilde \omega F_{\rm R} - G'_{\rm R}
-\tilde M_F(H F_{\rm L} + K G_{\rm L}) = 0
\ , \end{equation}
where $x$ is the dimensionless coordinate,
$\tilde \omega$ is the
dimensionless eigenvalue
$\tilde \omega = \omega /M_W$
and $\tilde M_F$ is the dimensionless fermion mass
$\tilde M_F= M_F/M_W$.

The eigenvalue problem Eqs.~(51)-(54) for the fermions
in a (bi)sphaleron-like background field
requires certain boundary conditions
for the fermion functions.
At the origin $G_{\rm L}(x)$ and $G_{\rm R}(x)$
are finite, while $F_{\rm L}(x)$ and $F_{\rm R}(x)$
vanish, at spatial infinity all functions vanish.

\subsection{\bf Fermionic bound states}

Previously we solved Eqs.~(51)-(54) in the background field
of the minimal energy sphaleron barrier for $M_H = M_W$,
i.~e.~$\epsilon=1/2$ [14].
Here we extend our analysis to larger values
of the Higgs mass, $M_H > 12 M_W$,
where the first bisphalerons are present.

For large Higgs masses the first bisphalerons
represent the top of the energy barrier.
Therefore we first consider the fermionic bound states
in the background field of these bisphalerons.
While the sphaleron always has a fermion zero mode,
independent of the fermion mass and of the Higgs mass,
the fermion eigenvalue in the bisphaleron background field
depends on both the fermion mass and on the Higgs mass.
In Fig.~5 the fermion eigenvalue $\omega$
is shown as a function of the
fermion mass for several values of the Higgs mass.
For small fermion masses we observe a critical value of the
fermion mass, $M_F^{\rm cr}$,
below which there is no bound state.
At the critical value $M_F^{\rm cr}$
the fermion mode reaches the
continuum, $\omega \rightarrow M_F$.
The critical value decreases with decreasing Higgs mass
and reaches zero at the first critical value of the Higgs mass,
$M_H^{\rm cr}=12 M_W$,
where the bisphalerons bifurcate from the sphaleron.
When the Higgs mass approaches this critical value,
the fermion eigenvalue approaches zero,
the eigenvalue of the sphaleron for all fermion masses.
In the bisphaleron background field
the fermion eigenvalue is zero
only at a particular value of the fermion mass.
This fermion mass is in the vicinity of 115 GeV,
almost independent of the Higgs mass,
increasing only very slightly with increasing Higgs mass.
The fermion eigenvalue in the background field of the parity conjugate
bisphaleron is $-\omega$.

Let us now turn to the fermion eigenvalue along the barrier.
First we consider the eigenvalue along the extremal energy path,
since this path is well defined and represents part of the
improved path, Eqs.~(38)-(43).
In Fig.~6a we present the fermion eigenvalue normalized
by the fermion mass for several values of the fermion mass
along the central part of the extremal path for
$\epsilon=100$.
The catastrophes present along the extremal path
are reflected in the behaviour of the fermion eigenvalue,
and there is a strong dependence on the fermion mass.
At the two critical points, where the energy reaches a spike
as a function of the Chern-Simons number,
the fermion eigenvalue bends around, having an infinite derivative,
provided that the fermion mass is large enough so that fermion
bound states still exist in this region of $N_{\rm CS}$.

For light fermions bound states exist only in a narrow
region around $N_{\rm CS}=1/2$, and we observe
three distinct branches, illustrated in Fig.~6a for $M_F=0.8$ GeV.
Moving along the extremal path, a first branch
starts from the positive continuum,
passes zero and dives into the negative continuum.
Then a second branch starts from the negative continuuum
passes zero at the sphaleron and dives into the positive
continuum. Finally a third branch starts from the positive continuum,
passes zero and dives into the negative continuum.
Thus three zero modes occur along the full extremal path,
but since these modes are encountered from alternating directions,
the total change in fermion number is the same as for a single
level crossing.

For heavier fermions, when bound states exist also at the
values of $N_{\rm CS}$, where the spikes are encountered,
the fermion eigenvalue starts from the positive continuum,
crosses zero, and bends backward and upward at the first spike.
Then it crosses zero again at the sphaleron,
reaches the second spike and bends forward and downward,
to cross zero for the third time and finally reach the negative
continuum. This is illustrated in Fig.~6a for
the fermion masses $M_F=8$ GeV and $M_F=80$ GeV.
For heavy fermions with $M_F>145$ GeV the fermion eigenvalue
crosses zero only once along the extremal path,
at the sphaleron, as illustrated in Fig.~6a for $M_F=160$ GeV.

The oscillating, catastrophic behaviour of the
fermion eigenvalues along the extremal path
for large Higgs masses is remedied for the improved path,
Eqs.~(38)-(43).
This is demonstrated in Fig.~6b for a fermion mass of $M_F=80$ GeV
and $\epsilon=100$, along the improved path presented
in Fig.~4b.
Here the eigenvalue decreases monotonically from the
positive continuum to the negative continuum
as a function of $N_{\rm CS}$, passing zero only once.

Let us now consider how the fermion eigenvalue
along the improved path depends on the fermion mass.
It is instructive to first inspect the critical values of the
fermion mass $M_F^{\rm cr}$, where the bound state
enters the continuum along the extremal path,
since to a large extend the extremal path coincides with
the improved path, Eqs.~(38)-(43).
The critical values of the fermion mass are illustrated in Fig.~7
for $\epsilon=100$.
While for small fermion masses fermion bound states exist only in the
vicinity of $N_{\rm CS}=1/2$,
fermion bound states exist with increasing fermion mass
along increasing parts of the barrier.
Reflecting again the catastophe,
we observe six critical values
below $M_F=6.8$ GeV and 2 critical values above $M_F=6.8$ GeV.

It is further instructive to observe
where along the extremal path
the zero modes occur for various fermion masses.
These fermion masses are shown in Fig.~8
for $\epsilon=100$.
For zero mass fermions the three zero modes
are encountered at $N_{\rm CS}=0.505$, $N_{\rm CS}=1/2$
and $N_{\rm CS}=0.495$ along the extremal path.
Moving along the improved path, Eqs.(38)-(43),
however, only one zero mode is encountered
for zero mass fermions, occurring at $N_{\rm CS}=0.505$
along the part $C_{\rm e}^+(N_{\rm CS})$ of the improved path,
which coincides with the extremal path.

Also for small fermion masses the fermion zero mode occurs
along the extremal part $C_{\rm e}^+(N_{\rm CS})$
of the improved path.
Fig.~8 gives the value of the fermion mass beyond which
the zero mode occurs along the intermediate part,
Eqs.~(42)-(43), of the improved path,
depending on the point of departure of the improved path
from the extremal path.
For the improved path presented in Fig.~4b this
fermion mass is $M_F=34$ GeV.
Since the path passes the bisphaleron,
the fermion zero mode is encountered before the bisphaleron
for fermions masses smaller than $M_F=116$ GeV
and beyond the bisphaleron
for fermions masses larger than $M_F=116$ GeV.
Of course the precise point along the path, where
the zero mode is encountered,
depends on the specific choice of the improved path,
except when the zero mode occurs in the bisphaleron
background field.

Analogous considerations hold for the parity
conjugate path, passing the parity conjugate bisphaleron.

\section{\bf Conclusions}

We have demonstrated the level crossing phenomenon
along the energy barrier between topologically distinct vacua
for large Higgs masses, when the barrier is no longer
symmetric.

We have employed various approximations in this study.
To keep spherical symmetry, we have neglected the finite
mixing angle and the mass splitting within the fermion doublet.
While the mixing angle dependence is small [21,22],
the effect of the fermion mass splitting on the fermion modes
certainly needs further investigation.
We have also only treated the valence fermion.
The inclusion of the Dirac sea has been studied by Diakonov et al. [16]
(for small Higgs masses),
showing only a small effect for light fermions, but an important
contribution for heavy fermions.
We have further neglected the back reaction of the fermions on the
barrier, which has been considered in [15]
(for small Higgs masses),
showing again that the effect is important only for heavy fermions,
when the barrier becomes asymmetric even for small Higgs masses.

For large Higgs masses,
when $M_H>12M_W$, there are two parity conjugate barriers,
and the configurations representing the top of these barriers
are the two conjugate first bisphalerons.
We have considered various paths over these barriers,
passing a bisphaleron. The extremal path [19]
is not satisfactory for these large Higgs masses,
since it is parity symmetric, passing besides both
bisphalerons also the sphaleron, with catastrophes
in the vicinity of the bisphalerons.
Therefore we have constructed an improved path
involving the single negative mode of the bisphaleron.
This improved path is smooth and culminates at the bisphaleron.

Solving the fermion eigenvalue equations in the background
field of the bisphaleron, we observe, that
the fermion eigenvalue depends on both the fermion mass
and on the Higgs mass. This is in contrast to the
sphaleron, which has a fermion zero mode for all fermion masses
and Higgs masses.
For a given Higgs mass there is only one particular value of the
fermion mass, where the bisphaleron supports a zero eigenvalue.
Interestingly, this fermion mass depends only very slightly
on the Higgs mass.

In the background field of the extremal path the fermion eigenvalue
reflects the catastrophic behaviour of the extremal path.
For small fermion masses, $M_F<145$ GeV,
there are three zero modes encountered
from alternating directions, while for larger fermion masses
there is only one zero mode encountered at the sphaleron.
In contrast, along the smooth improved path
the fermion eigenvalue is also smooth and only a single
zero mode is encountered for all fermion masses.
Interestingly, this zero mode occurs almost at $N_{\rm CS}=1/2$,
when the fermion mass tends to zero.

\vfill\eject

\vfill\eject
\section{Figure Captions}

\par

{\bf Fig.~1:}

 The sphaleron functions $f_A(x)$ and $K(x)$ are shown
 for three values of the Higgs mass,
 $M_H=20 M_W$ (dashed),
 $M_H=\sqrt{200} M_W$ (solid)
 and $M_H= M_W$ (dotted).

{\bf Fig.~2:}

 The bisphaleron functions $f_A(x)$, $f_B(x)$, $H(x)$ and $K(x)$
 are shown for two values of the Higgs mass,
 $M_H=20 M_W$ (dashed)
 and $M_H=\sqrt{200} M_W$ (solid).

{\bf Fig.~3:}

 The functions of the negative eigenmode of the bisphaleron
 $\eta_A(x)$, $\eta_B(x)$, $\eta_C(x)$, $\eta_H(x)$ and $\eta_K(x)$
 are shown for two values of the Higgs mass,
 $M_H=20 M_W$ (dashed)
 and $M_H=\sqrt{200} M_W$ (solid)
 with the normalization $\eta_H(0)=1$.

{\bf Fig.~4a:}

 The energy $E$ (in TeV) is shown as a function of the
 Chern-Simons number $N_{\rm CS}$ along the full extremal path
 from one vacuum to another for
 $M_H=20 M_W$ (solid) and
 $M_H= M_W$ (dashed).

{\bf Fig.~4b:}

 The energy $E$ (in TeV) is shown as a function of the
 Chern-Simons number $N_{\rm CS}$ in the critical region
 of the extremal path for
 $M_H=\sqrt{200} M_W$ (solid).
 Also shown is the energy for the configurations Eq.~(33)
 (dotted),
 and the energy along the improved path Eqs.~(38)-(43) (dashed).
 The improved path and the extemal path
 differ only in the intermediate region
 Eqs.~(42)-(43) but coincide at the bisphaleron.
 The crosses mark the bisphalerons and the plus signs the points
 of departure of the improved path from the extremal path.

{\bf Fig.~5:}

 The fermion eigenvalue $\omega$ (in GeV) in the background field
 of the bisphaleron is shown
 as a function of the fermion mass $M_F$ (in GeV)
 for three values of the Higgs mass,
 $M_H=20 M_W$ (dotted),
 $M_H=\sqrt{200} M_W$ (dashed)
 and $M_H=\sqrt{150} M_W$ (dot-dashed).
 Also shown are the critical value of the fermion mass
 $M_F^{\rm cr}$, where the fermion eigenvalue
 reaches the continuum (long-dashed),
 and the zero eigenvalue of the sphaleron (solid).

{\bf Fig.~6a:}

 The normalized fermion eigenvalue $\omega/M_F$
 is shown as a function of the Chern-Simons number $N_{\rm CS}$
 along the extremal path
 for the values of the fermion mass
 $M_F=0.8$ GeV, $M_F=8$ GeV, $M_F=80$ GeV and $M_F=160$ GeV
 for $M_H=\sqrt{200} M_W$ (solid),
 and for comparison for $M_F=0.8$ GeV, $M_F=8$ GeV and $M_F=80$ GeV
 for $M_H= M_W$ (dashed).

{\bf Fig.~6b:}

 The normalized fermion eigenvalue $\omega/M_F$
 is shown as a function of the Chern-Simons number $N_{\rm CS}$
 along the critical part of the extremal path
 for $M_H=\sqrt{200} M_W$
 for the fermion mass $M_F=80$ GeV (solid).
 Also shown is the normalized fermion eigenvalue $\omega/M_F$
 along the improved path Eqs.~(38)-(43) (dashed).
 Both paths differ only in the intermediate region
 Eqs.~(42)-(43) and cross at the bisphaleron.
 The crosses mark the bisphalerons and the plus signs the points
 of departure of the improved path from the extremal path.

{\bf Fig.~7:}

 The critical fermion mass $M_F^{\rm cr}$ (in GeV), at which
 the bound state enters the continuum
 for a given Chern-Simons number
 along the extremal path,
 is shown as a function of the Chern-Simons number $N_{\rm CS}$
 for $M_H=\sqrt{200} M_W$ (solid),
 and for comparison
 for $M_H= M_W$ (dashed).

{\bf Fig.~8:}

 The fermion mass $M_F$ (in GeV), at which
 the zero mode occurs
 for a given Chern-Simons number
 along the extremal path,
 is shown as a function of the Chern-Simons number $N_{\rm CS}$
 for $M_H=\sqrt{200} M_W$.
 The vertical line represents the zero mode of the sphaleron.

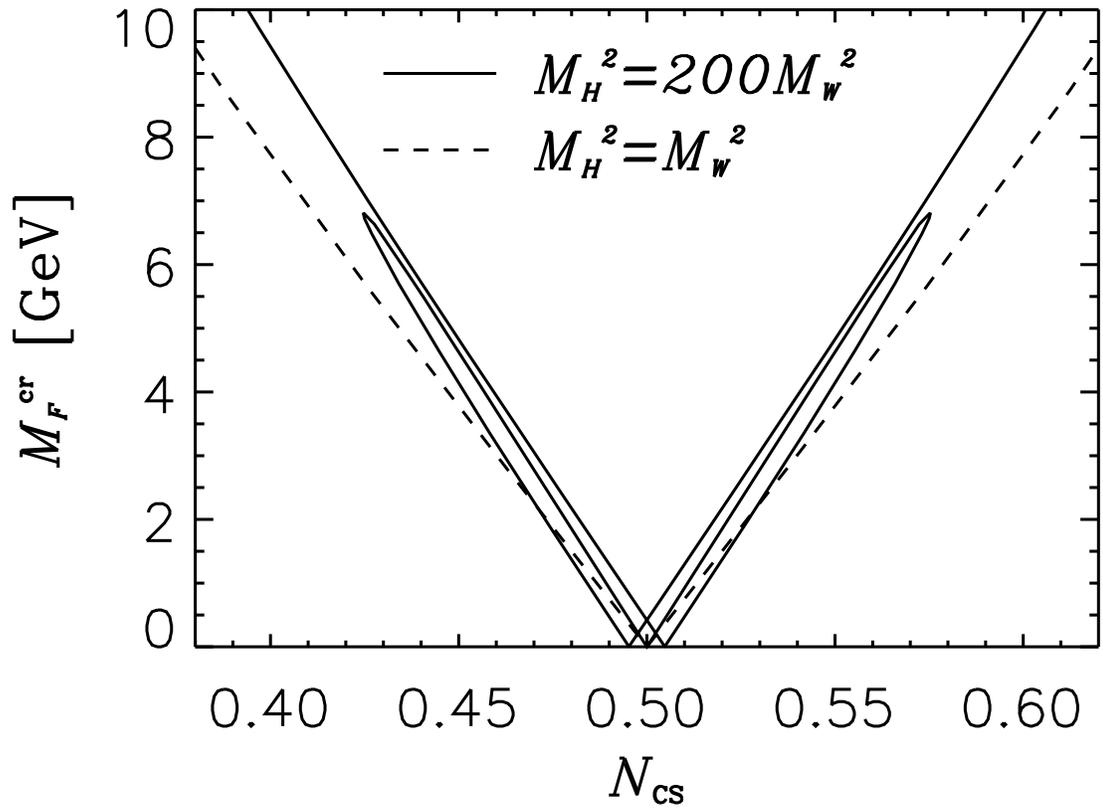
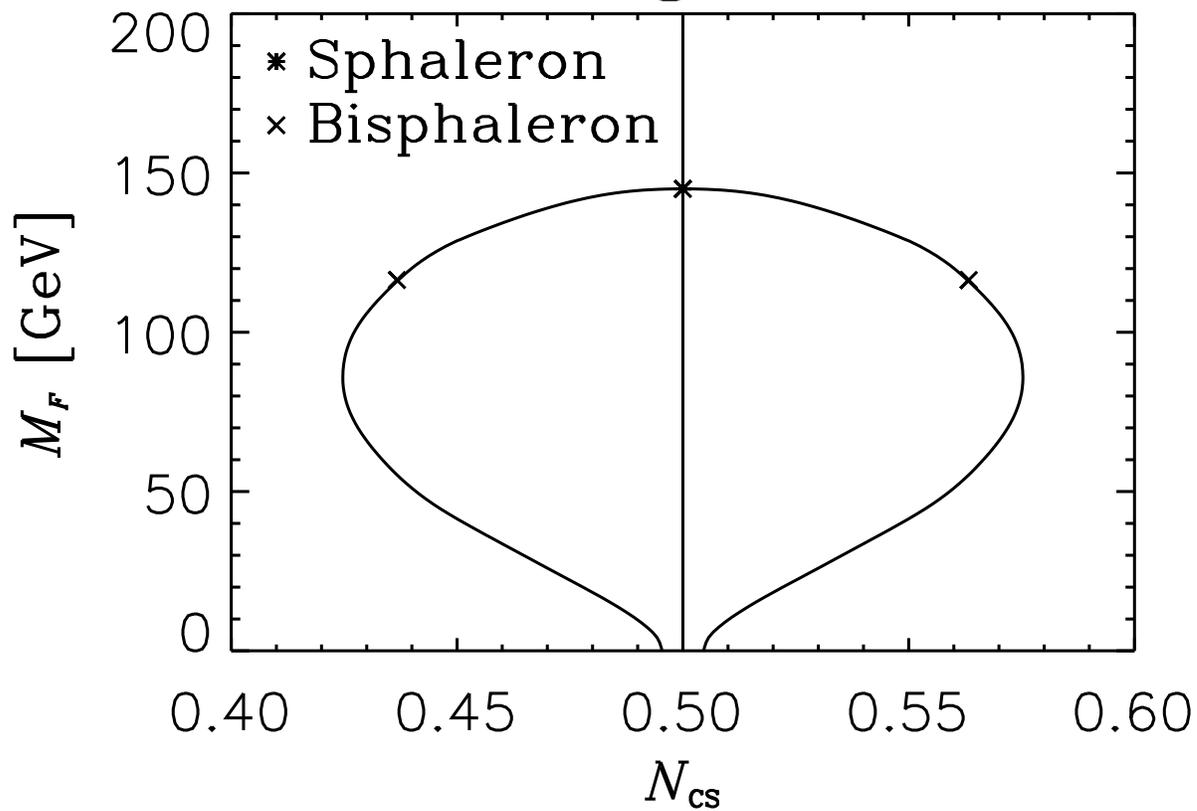
\end{document}